\documentclass[a4paper]{IEEEtran}
\usepackage{enumerate}
\usepackage{multirow,array}
\usepackage{cite}
\usepackage{graphicx}
\usepackage{psfrag}
\usepackage{caption}
\usepackage{subcaption}
\usepackage{url}
\usepackage{amsmath}
\usepackage{amsthm}
\usepackage{array}
\usepackage{amssymb}
\usepackage{amsfonts}
\usepackage{float}

\usepackage{algorithm}
\usepackage{algpseudocode}

\usepackage{color, colortbl}
\definecolor{Gray}{gray}{0.9}

\makeatletter
\renewcommand{\boxed}[1]{\text{\fboxsep=.2em\fbox{\m@th$\displaystyle#1$}}}
\makeatother
\newcommand{\C}{\cal}
\newcommand{\Lbarsq}{\overline{{\cal L}^2}}

\newcommand{\fq}{{\mathbb F}_q}
\newcommand{\gbinom}[2]{\begin{bmatrix}#1\\#2\end{bmatrix}_q}
\newtheorem*{conjecture*}{Conjecture}
\newtheorem{lemma}{Lemma}

\newtheorem{corollary}{Corollary}
\newtheorem{remark}{Remark}
\newtheorem{definition}{Definition}
\newtheorem{theorem}{Theorem}
\newtheorem{example}{Example}

\title{Low Subpacketization Coded Caching via Projective Geometry for Broadcast and D2D networks}
\begin{document}

\author{
\IEEEauthorblockN{Hari Hara Suthan Chittoor, Prasad Krishnan\\}
\IEEEauthorblockA{
Signal Processing and Communications Research Center,\\
International Institute of Information Technology, Hyderabad, India.\\
Email: \{hari.hara@research., prasad.krishnan@\}iiit.ac.in}
\vspace{-0.5cm}
}

\date{\today}
\maketitle
\thispagestyle{empty}	
\pagestyle{empty}
%%%%%%%%
\begin{abstract}
Coded caching was introduced as a technique of systematically exploiting locally available storage at the clients to increase the channel throughput via coded transmissions. Most known coded caching schemes in literature enable large gains in terms of the rate, however at the cost of subpacketization that is exponential in $K^{\frac{1}{r}}$ ($K$ being the number of clients, $r$ some positive integer). Building upon recent prior work for coded caching design via line graphs and finite-field projective geometries, we present a new scheme in this work which achieves a subexponential (in $K$) subpacketization of $q^{O((log_qK)^2)}$ and rate $\Theta\left(\frac{K}{(log_qK)^2}\right)$, for large $K$, and the cached fraction $\frac{M}{N}$ being upper bounded by a constant $\frac{2}{q^{\alpha-1}}$ (for some prime power $q$ and constant $\alpha \geq 2$) . Apart from this asymptotic improvement, we show that through some numerical comparisons that our present scheme has much lower subpacketization than previous comparable schemes, with some increase in the rate of the delivery scheme, for the same memory requirements. For instance, we obtain practically relevant subpacketization levels  such as $10^2 - 10^7$ for $10^2 - 10^4$ number of clients.  Leveraging prior results on adapting coded caching schemes for the error-free broadcast channel to device to device networks, we obtain a low-subpacketization scheme for D2D networks also, and give numerical comparison for the same with prior work. 

%Coded Caching is a promising solution to reduce the peak traffic in broadcast networks by prefetching the popular content close to end users and using coded transmissions. One of the chief issues of most coded caching schemes in literature is the issue of large \textit{subpacketization}, i.e., they require each file to be divided into a large number of subfiles. In this work, we present a coded caching scheme using line graphs of bipartite graphs in conjunction with projective geometries over finite fields. The presented scheme achieves a rate $\Theta(\frac{K}{\log_q{K}})$ ($K$ being the number of users, $q$ is some prime power) with  \textit{subexponential} subpacketization $q^{O((\log_q{K})^2)}$ when cached fraction is upper bounded by a constant ($\frac{M}{N}\leq \frac{1}{q^\alpha}$) for some positive integer $\alpha$). Compared to earlier schemes, the presented scheme has a lower subpacketization (albeit possessing a higher rate). We also present a new subpacketization dependent lower bound on the rate for caching schemes in which each subfile is cached in the same number of users. Compared to the previously known bounds, this bound seems to perform better for a range of parameters of the caching system.  

%\textit{Due to space restrictions, we give the full version of this paper in \cite{Chittoor2019LowSC} with all the proofs and further results. 

\end{abstract}
%%%%


%\vspace{-0.2cm}

\section{Introduction}
Next generation wireless networks (5G and beyond) present the challenges of serving clients via channels which are not traditional point-to-point communication channels. The study of efficient high throughput communication techniques for broadcast channels, interference channels, multiple access channels, device-to-device (D2D) communication, all obtain relevance in the present wireless communication scenario. The technique of utilizing local storage (which has become quite affordable, thanks to advances in hardware design), either on the client's device or in a nearby location, for aiding communication services on a broadcast channel was introduced formally in the landmark paper \cite{MaN}, under the title of \textit{Coded Caching}. In \cite{MaN}, it was shown that a combination of (a) carefully utilizing the local storage or \textit{cache} available to individual clients, and (b) coded transmissions during the delivery phase, brings tremendous gains in the rate of delivery of information of a broadcast channel. Following \cite{MaN}, generalized cache aided communication techniques have been presented for a number of channel models \cite{interferencemanagement,decentralizedcodedcaching,D2D,combinationnetworks}, and in each case has shown to provide gains in the information delivery rate. Specifically, the coded caching problem for D2D networks was considered in \cite{D2D}, and a caching cum coded delivery scheme that was inspired from \cite{MaN} was presented, which resulted similar rate advantages as \cite{MaN}. 

%The key performance challenges that next generation wireless networks (5G) face are low latency, high throughput and energy efficiency\cite{FTB}. Content delivery networks have been estimated to carry $72\%$ of the global internet traffic by $2022$ \cite{Cis}. 
%\textit{Coded caching} was proposed recently in a landmark paper by Maddah-Ali and Niesen \cite{MaN} and has emerged as an important tool to address major challenges of future communication networks. 
%Since its inception coded caching has proved as an efficient tool to trade-off expensive bandwidth with abundantly available and cost-effective memory 
% (due to advance in VLSI technology) at the user/network nodes.

The setup considered in \cite{MaN} consists of an error-free broadcast channel connecting a single server with $K$ clients or receivers. The server has $N$ same-sized files, which form the library of files. Each file is divided into $F$ equal-sized subfiles ($F$ is known as the \textit{subpacketization} parameter). Each client has a \textit{cache} that can store $MF$ subfiles (i.e., $\frac{M}{N}$ fraction of each file). According to the scheme presented in \cite{MaN} which takes place in two phases, the caches of the clients are populated by the subfiles during \textit{the caching phase} (which occurs during off-peak time periods), and during the \textit{demand phase} (occurring during peak-time periods) coded subfiles are transmitted to satisfy the client demands (each client demands one file in the demand phase). 
%The \textit{centralized coded caching} scheme of \cite{MaN} works in two phases. In the \textit{caching phase} (which occurs during off peak times) the cache of each client is populated with some $\frac{M}{N}$ fraction of each file in the server.
%This happens during off-peak times (when network resources are abundant). 
%In the \textit{delivery phase} (which happens during peak traffic times), the clients demand one file each from the server, to satisfy which the server sends coded transmissions. 
%This happens during peak times (when networks are congested).  
The \textit{rate} $(R)$ of such a coded caching scheme is defined as the ratio of the number of bits transmitted to the size of each file, which can be calculated as
\[
\text{Rate}~R = \small \frac{\text{Number of transmissions in the delivery phase}}{\text{Number of subfiles in a file}},
\] when each transmission is of the same size as the subfiles. 

The delivery scheme in \cite{MaN} consists of transmissions such that in each transmission $\gamma=1+\frac{MK}{N}$ clients are served. The parameter $\gamma$ is known as the \textit{global caching gain.} The rate achieved is $R=\frac{K(1-\frac{M}{N})}{\gamma}$. This rate was shown to be optimal for uncoded cache placement \cite{WTP}. The subpacketization $F$ of the scheme in \cite{MaN} is $F=\binom{K}{MK/N}$, which however becomes exponential in $K$ as $K$ grows (for constant $\frac{M}{N}$) and hence impractical even for tens of clients.

%The Ali-Niesen scheme in \cite{MaN} achieves $R=\frac{K(1-\frac{M}{N})}{\gamma}$, where $\gamma=1+\frac{KM}{N}$ is the \textit{global caching gain}, i.e., the number of users served by each transmission in the delivery scheme. This rate was shown to be optimal for uncoded cache placement \cite{WTP}. Further,  the subpacketization level used by the Ali-Niesen scheme to achieve this rate is $F=\binom{K}{\frac{KM}{N}}$.  Note that as $K$ grows large, $F\approx 2^{KH(M/N)}$, (for constant $\frac{M}{N}$, $H(.)$ being the binary entropy). This means that the files have to be extremely large  for even $50$-$100$ clients, making the Ali-Niesen scheme impractical for applications. 

%However, the subpacketization required by the Ali-Niesen scheme grows exponential in $K$, hence making the scheme impractical for applications.  
 Since then several new coded caching schemes with lower subpacketization have been constructed at the cost of increase in rate, or cache requirement, or the number of users (for instance, \cite{TaR} and \cite{YCTCPDA}). To the best of the author's knowledge, these constructions (and others in literature for which `large-$K$' behaviour can be derived) have subpacketization lesser than \cite{MaN} but still exponential in $K^{\frac{1}{r}}$ (for some positive integer $r$), while having larger rates compared to \cite{MaN}. In particular, the scheme in \cite{YCTCPDA} achieves global caching gain $\frac{MK}{N}$ with subpacketization exponential in $K$ with a much smaller exponent than \cite{MaN}, using a combinatorial structure called Placement Delivery Arrays (PDA). The issue of high subpacketization is carried over to the D2D problem also. Improved schemes with lower subpacketization (but higher rates) were also recently constructed for D2D networks in \cite{D2DPDAJan2019,hypercubeD2D}.

%mong these, an important construction was reported in \cite{YCTCPDA} via a construction known as \textit{Placement Delivery Arrays} (PDAs). The scheme of constructed in \cite{YCTCPDA} achieved a global caching gain of $\frac{MK}{N}$ (one less than that of \cite{MaN}), while improving the subpacketization by an exponential factor compared to \cite{MaN}. However, in this construction, as well as in most others in literature, the subpacketization required for the caching schemes continues to be exponential in $K^{\frac{1}{r}}$ (for some positive integer $r$) to the best of our knowledge.  

In this direction of research, a line graph based coded caching scheme was introduced and developed in \cite{haribhavanaprasad,PK} (co-authored by a subset of the current authors) to construct one of the few explicitly known \textit{subexponential} (in $K$) subpacketization schemes. Tools from projective geometries over finite fields were used for this purpose. However the scheme of \cite{PK} required a large cache requirement to obtain low subpacketization. This issue was rectified in \cite{haribhavanaprasad}. The scheme in \cite{haribhavanaprasad} achieves a rate of $\Theta(\frac{K}{\log_q{K}})$ ($K$ being the number of users, $q$ is some prime power) with  subexponential subpacketization $q^{O((\log_q{K})^2)}$ when cached fraction is upper bounded by a constant ($\frac{M}{N}\leq \frac{1}{q^\alpha}$) for some positive integer $\alpha$). 

%Recently a bipartite graph framework for coded caching was presented in \cite{YTCC}. In this work the authors modeled the users as left vertices and subfiles as right vertices of a bipartite graph. The edges of the bipartite graph represent the uncached subfiles. A \textit{strong edge coloring} of this bipartite graph gives the transmission scheme. [PRASAD Sir] 
%Recently, a line graph based approach to coded caching was introduced in \cite{PK}. %In \cite{PK}, a caching scheme was shown to be equivalent to a line graph of a bipartite graph.  A clique cover of the square of the complement of the line graph can then be used to construct a delivery scheme. 
%Using this framework, a construction for a caching scheme was given via a projective geometry over a finite field. The scheme presented in \cite{PK} achieves a constant rate with subpacketization subexponential in $K$ $\left(F=q^{O((log_q K)^2)}\right)$ for some prime power $q$). However the drawback of this scheme is that the uncached fraction of each file has to be large $\left((1-\frac{M}{N})=\Theta(\frac{1}{\sqrt{K}})\right)$. We remedy this drawback (to some extent) in this work. 

%In this authors modeled the coded caching problem as line graph of bipartite caching graph. The transmissions are given as the clique cover of the complement of square of the caching line graph.

In the present work, we go further than the scheme of \cite{haribhavanaprasad}. The tools remain the same; we use a line graph based technique combined with projective geometries over finite fields. The contributions and organization of the current work are as follows. After briefly going over the line graph coded caching approach in \cite{PK} (Section \ref{review}), we present our new scheme in Section \ref{ourscheme}. In Section \ref{asymptotics}, we show that for large $K$ and the cached fraction $\frac{M}{N}\leq \frac{2}{q^{\alpha-1}}$ (for some constant $\alpha \geq 2$), we show that our scheme achieves rate $\Theta(\frac{K}{(log_qK)^2})$ and subpacketization $q^{O((log_qK)^2)}$ for large $K$, thus improving upon \cite{haribhavanaprasad}. Further we also compare in Table \ref{tab1}, by giving some numerical values to our scheme's parameters with those of \cite{YCTCPDA} and \cite{haribhavanaprasad}, and show that the subpacketization achieved is several orders of magnitude lesser compared to \cite{haribhavanaprasad} (which itself is orders of magnitude less than \cite{YCTCPDA}). However the rate (equivalently, the gain) of our present scheme can be few orders of magnitude greater (equivalently, lesser) than \cite{YCTCPDA} and roughly the same as \cite{haribhavanaprasad}. Finally, in Section \ref{d2d}, we extend the present scheme for the error-free broadcast channel to D2D networks, utilizing a result from \cite{PDAapplications}. This results in a D2D coded caching scheme with lower subpacketization than some known schemes before. In Table \ref{tab2} we perform a numerical comparison of the new D2D scheme with those of \cite{hypercubeD2D,D2D}.

%The contributions and organization of this paper are as follows. In Section \ref{review}, we review the line graph based coded caching scheme proposed in \cite{PK}, while refining it slightly for our purposes. In Section \ref{lowerbound}, we propose a new lower bound for the optimal rate $R^*$ given parameters $K, F,$ and $\frac{M}{N}$, for the caching schemes in which each subfile is cached in the same number of users, which is a property satisfied by all known centralized caching designs in literature (to the best of our knowledge). Using some numerical examples, we see that this lower bound performs better (for a range of parameters) compared to the previously known bounds in \cite{WTP,cheng2017coded}. In Section \ref{ourscheme}, we present a new coded caching scheme using projective geometries over finite fields in the line graph framework of \cite{PK}. In Section \ref{asymptotics} we give the asymptotic analysis of the scheme proposed. We show that the scheme achieves a rate $R=\Theta\left(\frac{K}{log_q K}\right)$ ($q$ being a prime power) for a constant cache requirement, which can be extremely small ($\frac{M}{N}\leq \frac{1}{q^{\alpha}}$, for some constant positive integer $\alpha$). The subpacketization achieved is subexponential in $K$, $F=q^{O((\log_q{K})^2)}$. We provide a table in Section \ref{asymptotics} which compares the parameters of our scheme to that of \cite{YCTCPDA}.  %Finally in Section VI we conclude the paper and give possible future directions.

\textit{Notations and Terminology:} $\mathbb{Z}^{+}$ denotes the set of positive integers. 
We denote the set $\{1,\hdots,n\}$ by $[n]$ for some positive integer $n$. For sets $A,B$, the set of elements in $A$ but not in $B$ is denoted by $A\backslash B$. The finite field with $q$ elements is $\fq$. The dimension of a vector space $V$ over $\fq$ is given as $dim(V)$. For two subspaces $V,W$, their subspace sum is denoted by $V+W$. Note that $V+W=V\oplus W$ (the direct sum) if $V\cap W=\phi$. The span of two vectors $\mathbf{v_1},\mathbf{v_2}\in V$, is represented as $span(\mathbf{v_1},\mathbf{v_2})$.
We give some basic definitions in graph theory. The sets $V(G), E(G)$ denote vertex set and edge set of a (simple undirected) graph $G$ respectively, where $E(G)\subseteq \left\{\{u,v\}:u,v\in V(G), u\neq v\right\}$.
%The neighbourhood of a vertex $u\in V(G)$ is given as $\mathcal{N}(u)=\{v\in V(G): \{u,v\}\in E(G)\}$. 
The square of a graph $G$ is a graph $G^2$ having $V(G^2)=V(G)$ and an edge $\{u,v\}\in E(G^2)$ if and only if either $\{u,v\}\in E(G)$ or there exists some $v_1\in V(G)$ such that $\{u,v_1\},\{v_1,v\}\in E(G)$. The complement of a graph $G$ is denoted as $\overline{G}$. A set $H\subseteq V(G)$ is called a clique of $G$ if every two distinct vertices in $H$ are adjacent to each other. A single vertex is also considered as a clique by definition. A clique cover of $G$ is a collection of disjoint cliques such that each vertex appears in precisely one clique. 
%A \textit{bipartite graph} $B$ is a graph whose vertices can be partitioned into two independent sets (called left and right vertices of $B$) such that edges exist only between left and right vertices. 
%A bipartite graph is (left or right) regular if the degree of each vertex on (left or right) is same throughout (left or right) partition. A bipartite graph is bi-regular if it is both left regular and right regular. For more information on graph theory the reader is referred to \cite{Die}.

%%%%%%
\section{The line graph based coded caching of \cite{PK} and its relation to PDAs}
\label{review}
Consider a coded caching system consisting of a server with files $\{W_i:i\in [N]\}$. Let $\mathcal{K}$ be any set such that $|\mathcal{K}|=K$. We shall use $\mathcal{K}$ to indicate the set of $K$ users. Let $\mathcal{F}$ be any set such that $|\mathcal{F}|=F$. The subfiles of a file $W_i$ are denoted by $W_{i,f}$ where $f\in \mathcal{F}$ and $W_{i,f}$ takes values in some Abelian group. %Here we consider \textit{symmetric caching}, i.e., for any $f\in \mathcal{F}$, either a user caches $W_{i,f}, \forall i\in[N]$ or the user does not cache $W_{i,f}$  for any $i\in[N]$. 

%Any symmetric caching scheme can be represented as an equivalent $D$-left regular \textit{bipartite caching graph} $B(K,D,F)$ with left vertices being $\mathcal{K}$ and right vertices being $\mathcal{F}$, and the uncached fraction $1-\frac{M}{N}=\frac{D}{F}$. The uncached subfiles are identified by the edges of $B$ i.e, for $k\in\mathcal{K}, f\in\mathcal{F}$ an edge $\{k,f\}\in E(B)$ if and only if the subfiles $W_{i,f}, \forall i\in[N]$ are not present in the cache of user $k$. This bipartite coded caching setup was given in \cite{YTCC}. 

%In \cite{PK}, a line graph based framework was proposed  to study the coded caching problem. The \textit{line graph} ${\C L}(G)$ (or simply, ${\cal L}$) of an undirected graph $G$ is a graph in which the vertex set $V({\C L}(G))$ is the edge set $E(G)$ of $G$, and two vertices of $V({\C L}(G))$ are adjacent if and only if they share a common vertex in $G$. The caching scheme was captured via a line graph ${\mathcal{L}}$ of a bipartite caching graph and the delivery scheme was obtained as a clique cover of complement of the square of line graph denoted by ${\Lbarsq}$. The following lemma proved in \cite{PK} presents the conditions under which an arbitrary graph is a line graph of a left regular bipartite graph. This enables us to construct a line graph which corresponds to a coded caching scheme.

In \cite{PK}, a line graph based framework was proposed  to study the coded caching problem, which we now describe.

\begin{definition}(Line graph, $(c,d)$-caching line graph) \cite{PK}
\label{line graph and (c,d) graph}
A graph ${\cal L}$ consisting of $KD$ vertices (for some $D\in \mathbb{Z}^+$), such that ${\cal V}({\cal L})\subseteq {\C K} \times {\C F}$, (for some sets $\mathcal{K,F}$ such that $|\mathcal{K}|=K,|\mathcal{F}|=F$) is said to be a caching line graph (or simply, a line graph) if
\begin{enumerate}
    \item[P1:] The set of vertices ${\C U}_k=\{(k,f) \in V({\C L}):f\in {\C F}\}$ forms a clique of size $D$, for each $k\in {\cal K}.$ (We refer to these cliques as the \textit{user cliques}).
    \item[P2:] The set of vertices ${\C S}_f=\{(k,f) \in V({\C L}):k\in {\C K}\}$ forms a clique of size $c$ (for some fixed $c \in \mathbb{Z}^+$), for each $f\in{\cal F}.$ (We refer to these cliques as the \textit{subfile cliques}).
    \item[P3:] Each edge in ${\C L}$ lies between vertices within a user clique or within a subfile clique.
    %\item 
\end{enumerate}
When there is a clique cover of $\Lbarsq$ (the complement of the square of ${\C L}$) consisting of disjoint $d$-sized cliques, then the line graph $\C L$ is called as a $(c,d)$\textit{-caching line graph}.

\end{definition}

By the above conditions P1-P3, it holds that $\bigcup\limits_{k\in {\cal K}}{\C U}_k=\bigcup\limits_{f\in {\C F}}{\C S}_f=V({\C L})$ (these unions being disjoint). %Furthermore, we should hence have $|{\C U}_k\cap {\C S}_f|\leq 1, \forall (k,f) \in {\C K}\times {\C F}.$ 
Therefore we can write  $V({\cal L})=\{(k,f)\in{\cal K}\times{\cal F}:{\cal U}_k\cap{\cal S}_f\neq \phi\}$. %With this notation, we have ${\C U}_k=\{(k,f) \in V({\C L}):f\in {\C F}\}$ and ${\C S}_f=\{(k,f) \in V({\C L}):k\in {\C K}\}$. 
Furthermore, it follows that $E({\cal L})=\{\{(k,f),(k',f')\}\subset V({\cal L}): k=k' \text{ or } f=f' \text{ but not both }\}$. 

It was shown in \cite{PK} (refer Section IV Proposition $1$ of \cite{PK}) that such a line graph ${\cal L}$ corresponds to a caching system in which there are $K$ users (indexed by ${\C K}$) and $F$ subfiles (indexed by ${\C F}$), where the $k^{th}$ user caches subfiles $\{W_{i,f}\in {\C F}:\forall i\in[N]\}$ if $(k,f)\notin V({\C L})$ and does not cache them otherwise. We therefore have that each user does not cache $D$ subfiles of each file, and hence the \textit{uncached fraction} is $1-\frac{M}{N}=\frac{D}{F}$. Further each subfile of any file is not cached in $c$ of the $K$ users.

It is also shown in \cite{PK} that for the caching phase as defined by the line graph $\C L$,  a delivery scheme is given by a clique cover of $\Lbarsq$. 
\begin{remark}
\label{cdremark}
For a $(c,d)$-caching line graph, it is shown in Theorem $2$ of \cite{PK} that the parameters of the caching and delivery scheme come out naturally, with $F=\frac{KD}{c}$ (and thus the uncached fraction being $1-\frac{M}{N}=\frac{c}{K}$). Further the $d$-sized cliques of $\Lbarsq$ result in a delivery scheme with rate $R=\frac{c}{d}$. This is illustrated in the following example.
\end{remark}

\begin{figure}
  \centering
                \includegraphics[height=1.1 in]{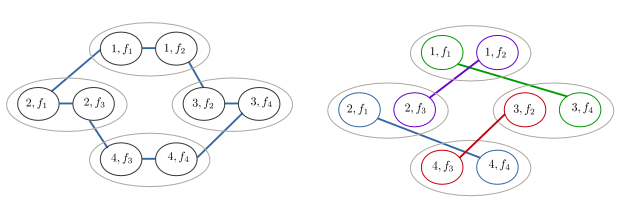}
                \caption{\small The graph $\mathcal{L}$ (left) and $\Lbarsq$ (right) corresponding to the Example \ref{example}. The user cliques in $\C L$ are placed within the ellipses, while the subfile cliques are the vertices across two user cliques connected by an edge. The cliques of $\Lbarsq$ are vertices of the same color, also connected by each edge.}
\label{fig:L and Lbarsq}
       \end{figure}

\begin{example}\label{example}
Consider a coded caching system defined by the graph $\cal L$ as shown in the left of Fig. \ref{fig:L and Lbarsq}. This graph $\cal L$ corresponds to a coded caching setup with $K=4$ (since it has 4 user cliques, each of size $D=2$) and $F=4$ (since it has 4 subfile cliques, each of size $c=2$). Thus, there are $4$ users, each user does not cache $D=2$ subfiles (for instance, user $1$ does not cache the subfiles indexed by $f_1$ and $f_2$ and caches $f_3,f_4$). Each subfile is cached in $c=2$ users (for instance, subfile $f_2$ is cached in users $1,3$). The cached fraction is $\frac{M}{N}=1-\frac{D}{F}=\frac{1}{2}$. The graph $\Lbarsq$ is shown on the right with 4  cliques, indicated by vertices of a same color. Corresponding to each clique of $\Lbarsq$, there is one transmission in the delivery scheme. For instance, corresponding to the clique $\{(1,f_2),(2,f_3)\}$, there is a transmission $W_{d_2,f_3}+W_{d_1,f_2}$, where $W_{d_k}$ signifies the file demanded by client $k$. Note that this enables the client $1$ to decode $W_{d_1,f_2}$ and client $2$ to decode $W_{d_2,f_3}$, as they each have the other subfile in their cache. Similarly, the entire set of $4$ transmissions will enable decoding of all the missing subfiles at all clients.

%with $K=4,F=4,D=2$ (i.e. $\frac{M}{N}=1-\frac{D}{F}=$. Therefore $\frac{M}{N}=1-\frac{c}{d}=\frac{1}{2}$. The $(c=2,d=2)$ caching line graph($\mathcal{L}$) and complement of square of line graph($\overline{\mathcal{L}^2}$) is shown in Fig \ref{fig:L and Lbarsq}. There are $4$ cliques present in $\Lbarsq$. The corresponding transmissions are $W_{d_{1,f_1}}+W_{d_{3,f_4}}$, $ W_{d_{1,f_2}}+W_{d_{2,f_3}}$, $W_{d_{2,f_1}}+W_{d_{4,f_4}}$ and $W_{d_{3,f_2}}+W_{d_{4,f_3}}$.
\end{example}
%%%

In the forthcoming sections, we construct a new caching line graph based scheme using  projective geometry over finite fields, building on the results of \cite{PK,haribhavanaprasad}, and show that these results outperform prior known schemes in terms of the subpacketization $F$, while trading it off with some increase in the rate of the delivery scheme. Towards that end, we now recall a following structural lemma (which will be used in the next section) proved in \cite{haribhavanaprasad} which gives the conditions under which an edge exists in ${\Lbarsq}$. 
\begin{lemma} \cite{haribhavanaprasad}
\label{edge in compliment of square}
Let  $(k_1,f_1),(k_2,f_2) \in V(\mathcal{{\C L}})$.
The edge $\{(k_1,f_1),(k_2,f_2)\}\in E({\Lbarsq})$ 
if and only if  $k_1\neq k_2, f_1\neq f_2$ and $(k_1,f_2)\notin V(\mathcal{L}), (k_2,f_1)\notin V(\mathcal{L})$.
\end{lemma}

Now we recall the definition of \textit{placement delivery array}(PDA) presented in \cite{YCTCPDA}.
\begin{definition}[Placement delivery array \cite{YCTCPDA}] 
\label{PDA definition}
For positive integers $K,F,Z$ and $S$ an $F\times K$ array $\boldsymbol{A}= [a_{j.k}],j\in [F],k\in [K]$, composed of a specific symbol ``$*$" and $S$ integers $1,\cdots,S,$ is called a $(K,F,Z,S)$ placement delivery array (PDA), if it satisfies the following conditions:
\begin{enumerate}
    \item[C1.] The symbol ``$*$" appears $Z$ times in each column.
    \item[C2.] Each integer occurs at least once in the array.
    \item[C3.] For any two distinct entries $a_{j_1,k_1}$ and $a_{j_2,k_2}$ we have $a_{j_1,k_1}=a_{j_2,k_2} =s$, an integer, only if
    \begin{enumerate}
        \item[1.] $j_1\neq j_2,k_1\neq k_2,$ i.e., they lie in distinct rows and distinct columns; and
        \item[2.] $a_{j_1,k_2}=a_{j_2,k_1} =*$.
    \end{enumerate}
\end{enumerate}

% i.e., the $2\times 2$ sub-array formed by rows $j_1,j_2$ and columns $k_1,k_2$ must be of the following form $\left[\begin{array}{cc}
%             s & * \\
%             * & s
%         \end{array}\right]$ or$\left[\begin{array}{cc}
%             * & s \\
%             s & *
%         \end{array}\right]$.

If each integer $s\in [S]$ occurs exactly $g$ times, $\boldsymbol{A}$ is called a regular $g-(K,F,Z,S)$ PDA, or $g$-PDA for short.
\end{definition}

Most known coded caching schemes in literature correspond to PDAs. 
We now show that any $(c,d)-$ caching line graph $(d\geq 2)$ is equivalent to a PDA.

\begin{lemma}
\label{line graph pda connection}
$\mathcal{L}$ is a $(c,d)$-caching line graph $($when $d\geq 2$ and there is a partition of $V(\mathcal{L})$ with $K$ cliques of size $D$ each$)$ if and only if there exist a $d-(K,F=\frac{KD}{c},Z=F-D,S=\frac{KD}{d})$ regular PDA.
\end{lemma}

\begin{IEEEproof}
We will prove the only if part. Let $\mathcal{L}$ be a caching line graph as given in the lemma statement. 
From the condition P2 of Definition \ref{line graph and (c,d) graph}, we have that the $c$-sized disjoint cliques of $\mathcal{L}$ partition $V(\mathcal{L})$. Since $|V(\mathcal{L})|=KD$, thus $\frac{KD}{c}$ is an integer. Also we know that there is a clique cover of $\Lbarsq$ consisting of $d$-sized disjoint cliques. As $V(\Lbarsq)=V(\mathcal{L})$, the set $V(\mathcal{L})$ can be partitioned into $\frac{KD}{d}$ number of $d$-sized cliques $\{C_i:i\in [\frac{KD}{d}]\}$ of $\Lbarsq$. It is clear that $|C_i|=d, \forall i\in [\frac{KD}{d}] $.
Let $F=\frac{KD}{c}$,  $Z=F-D$ and $S=\frac{KD}{d}$.
Now consider an array $\mathbf{A} = [a_{f,k}], f\in [F],  k\in [K]$. So $\mathbf{A}$ is a $F\times K$ array such that rows represent subfile cliques and columns represent user cliques. The entries of $\mathbf{A}$ are defined as follows
\begin{equation*}
    a_{f,k}=\begin{cases}
    * \textit{ if } (k,f)\notin V(\mathcal{L}) \\
    s \textit{ if } (k,f) \in C_s \textit{ for some } s\in [\frac{KD}{d}]
    \end{cases}
\end{equation*}
Now we will check the conditions C1-C3 of Definition \ref{PDA definition}.
\begin{enumerate}
    \item[C1.] Consider an arbitrary $k\in [K]$. By condition P1 of Definition \ref{line graph and (c,d) graph}, $|\{f\in [F]: (k,f) \notin V(\mathcal{L})\}|=F-D=Z$. Therefore ``$*$" appears $Z$ times in each column of $\mathbf{A}$.
    
    \item[C2.] From the definition of $\mathbf{A}$, it is clear that each integer $s\in [S]$ occurs at least once in the array.
    
    \item[C3.] Consider $a_{f_1,k_1},a_{f_2,k_2}$ such that $(k_1,f_1),(k_2,f_2)\in C_s$ for some $s\in [\frac{KD}{d}]$. From Lemma \ref{edge in compliment of square}, it is easy to see that $f_1\neq f_2, k_1\neq k_2$ and $(k_1,f_2),(k_2,f_1)\notin V(\mathcal{L})$. Therefore $a_{f_1,k_2}=a_{f_2,k_1}=*$.
\end{enumerate}

Therefore $\mathbf{A}$ satisfies all the conditions of Definition \ref{PDA definition}.
Hence $\mathbf{A}$ is a $d-(K,F,Z,S)$ PDA. The proof of if part follows similarly.
\end{IEEEproof}

\section{A new projective geometry based scheme}
\label{ourscheme}
Towards presenting our new scheme, we first review some basic concepts from projective geometry.
\vspace{-0.3cm}
\subsection{Review of projective geometries over finite fields \cite{hirschfeld1998projective}}
Let $k,q\in \mathbb{Z}^+$ such that $q$ is a prime power. Let $\fq^k$ be a $k$-dim (we use ``dim'' for dimensional) vector space over a finite field $\fq$. Consider an equivalence relation on $\fq^k\setminus \{\boldsymbol{0}\}$(where $\boldsymbol{0}$ represents the zero vector) whose equivalence classes are $1$-dim subspaces (without $\boldsymbol{0}
$) of $\fq^k$. The $(k-1)$-dim \textit{projective space} over $\fq$ is denoted by $PG_q(k-1)$ and is defined as the set of these equivalence classes. For $m\in [k]$, let $PG_q(k-1,m-1)$ denote the set of all $m$-dim subspaces of $\fq^k$.
It is known that (Chapter $3$ in \cite{hirschfeld1998projective}) $|PG_q(k-1,m-1)|$ is equal to the \textit{q-binomial coefficient} $\gbinom{k}{m}$, where
$
\begin{bmatrix}k\\m\end{bmatrix}_q
%=\frac{(q^k-1)(q^k-q)\hdots (q^k-q^{m-1})}{(q^m-1)(q^{m}-q)\hdots (q^m-%q^{m-1})}
=\frac{(q^k-1)\hdots(q^{k-m+1}-1)}{(q^m-1)\hdots(q-1)}
$ (where $k\geq m$).
In fact, $\gbinom{k}{m}$ gives the number of $m$-dim subspaces of any $k$-dim vector space over $\fq$. Further, by definition, $\gbinom{k}{0}=1.$
%In any Gaussian binomial coefficient $\gbinom{a}{b}$ given in this paper we assume that $a,b\in \mathbb{Z}^+$ and $1\leq b\leq a.$

Let $\mathbb{T} \triangleq \{T : T \in PG_q(k-1,0) \}$. Let $\theta(k)$ denotes the number of distinct $1$-dim subspaces of $\mathbb{F}_q^k$. Therefore $\theta (k) = |\mathbb{T}|=\gbinom{k}{1}= \frac{q^k -1}{q-1}$.  

% The following result is known from \cite{hirschfeld1998projective}.
% %
% \begin{lemma}[Chapter 3 in \cite{hirschfeld1998projective}] \label{no of subspaces}
% Consider a $k$-dim vector space $\mathbb{F}_q^k$.
% \begin{itemize}
%     \item 
    
%     \item Let $1\leq r,s,l <k$. Then the number of $r$-dim subspaces intersecting a fixed $s$-dim subspace in a fixed $l$-dim subspace is $q^{(r-l)(s-l)}\gbinom{k-s}{r-l}$.
% \end{itemize}

% \begin{itemize}
%     \item[(a2)] The number of $r$-dimensional subspaces intersecting a fixed $s$-dimensional subspace in some $l$-dimensional subspace are $q^{(r-l)(s-l)}\gbinom{n-s}{r-l} \gbinom{s}{t}$.
    
%     \item[(a3)] The number of ordered pairs $(s\textit{-dimensional subspcae }, r\textit{-dimensional subspace})$ intersecting in a fixed $l$-dimensional subspace are 
%     $q^{(r-l)(s-l)}\gbinom{n-s}{r-l} \gbinom{n-l}{n-s}$.
    
%     \item[(a4)] The number of ordered pairs $(s\textit{-dimensional subspcae }, r\textit{-dimensional subspace})$ intersecting in some $l$-dimensional subspace are 
%     $q^{(r-l)(s-l)}\gbinom{n-s}{r-l} \gbinom{n-l}{n-s} \gbinom{n}{l}$.

% \end{itemize}

% (Here all subspaces are subspaces of $\mathbb{F}_q^n$)

% \end{lemma}
The following lemma and corollary will be used repeatedly in this paper.
\begin{lemma} \label{no of sets of LI 1D spaces}
Let $k,a,b \in \mathbb{Z}^+$ such that $1\leq a+b\leq k$. Consider a $k$-dim vector space $V$ over $\fq$ and a fixed $a$-dim subspace $A$ of $V$. The number of distinct (un-ordered) $b$-sized sets $\{T_1,T_2,\cdots,T_b\}$ such that $T_i\in \mathbb{T}, \forall i\in[b]$ and $A\oplus T_1 \oplus T_2 \oplus \cdots \oplus T_b \in PG_q(k-1,a+b-1)$ is 
$\frac{\prod\limits_{i=0}^{b-1}(\theta(k)-\theta(a+i))}{b!}$.
\end{lemma}

\begin{IEEEproof}
First we find the number of $T_1\in \mathbb{T}$ such that $A\oplus T_1$ is a $(a+1)$-dim subspace of $V$. To pick such a $T_1$ we define, $T_1= span(\mathbf{t_1})$ for some $\mathbf{t_1}\in V \setminus A$. Such a $\mathbf{t_1}$ can be picked in $(q^k-q^{a})$ ways. However for one such fixed $\mathbf{t_1}$,
there exist $(q-1)$ number of $\mathbf{t_1'} (=\beta \mathbf{t_1}$, where $\beta \in \fq\backslash 0$) such that $span(\mathbf{t_1})=span(\mathbf{t_1'})=T_1$. Thus the required number of unique $T_1\in \mathbb{T}$ is $\frac{q^k-q^{a}}{q-1}=\theta(k)-\theta(a)$. Similarly for every such $T_1$ we can select $T_2$ with the condition that $A\oplus T_1 \oplus T_2$ is $(a+2)$-dim subspace of $V$ in $(\theta(k)-\theta(a+1))$ ways. So the number of distinct ordered sets $\{T_1,T_2\}$ is $(\theta(k)-\theta(a))(\theta(k)-\theta(a+1))$. By induction the number of distinct ordered sets $\{T_1,T_2, \cdots, T_b\}$ is $\prod\limits_{i=0}^{b-1}(\theta(k)-\theta(a+i))$.
We know that the number of permutations of a $b$-sized set is $b!$. Therefore the number of distinct (un-ordered) sets satisfying the required conditions is $\frac{\prod\limits_{i=0}^{b-1}(\theta(k)-\theta(a+i))}{b!}$. This completes the proof.
\end{IEEEproof}

\begin{corollary} \label{no of 1D spaces outside a hper space}
Consider two subspaces $A,A'$ of a $k$-dim vector space $V$ over $\fq$ such that $A'\subseteq A, dim(A)=a, dim(A')=a-1$. The number of distinct $T\in \mathbb{T}$ such that $A'\oplus T=A$ is $q^{a-1}$.
\end{corollary}
% \begin{IEEEproof}
% $\theta(a)-\theta(a-1)=\frac{q^a-q^{a-1}}{q-1}=q^{a-1}$.
% \end{IEEEproof}
We now proceed to construct a caching line graph using projective geometry.
% \vspace{-0.8cm}
\subsection{A new caching line graph using projective geometry} \label{our scheme B subsection}
Consider $k,m,t \in \mathbb{Z}^+$ such that $m+t+2\leq k$. Consider a $k$-dim vector space $\fq^{k}$. Let $W$ be a fixed $(t-1)$-dim subspace of $\fq^{k}$. 
Consider the following sets of subspaces, where each such subspace contains $W$.
\begin{align*}
    \mathbb{V} &\triangleq \{V \in PG_q(k-1,t-1): W\subseteq V \}.\\
    \mathbb{R} &\triangleq \{R \in PG_q(k-1,t): W\subseteq R \}.\\
    \mathbb{S} & \triangleq \{S \in PG_q(k-1,m+t-1): W\subseteq S\}.\\
    \mathbb{U} & \triangleq \{U \in PG_q(k-1,m+t+1): W\subseteq U\}.
\end{align*}

Now, consider the following sets, which are used to present our line graph and the corresponding coded caching scheme.

\begin{align}
    \label{eqn X}
    \mathbb{X} & \triangleq \left\{\{V_1,V_2\}: V_1,V_2 \in \mathbb{V}, V_1 +V_2\in \mathbb{R} \right\}.
\end{align}
\begin{align}
    \label{eqn Y}
    \mathbb{Y} & \triangleq \left\{\{V_1,V_2,\cdots,V_{m+1}\}: \forall V_i \in \mathbb{V}, \sum\limits_{i=1}^{m+1}V_i \in \mathbb{S} \right\}. \\
    \label{eqn Z}
    \mathbb{Z} & \triangleq \left\{\{V_1,V_2,\cdots,V_{m+3}\}: \forall V_i \in \mathbb{V}, \sum\limits_{i=1}^{m+3}V_i \in \mathbb{U}\right\}.
\end{align}

To construct a caching line graph $\C L$, we need to satisfy the conditions P1-P3 in Definition \ref{line graph and (c,d) graph}. Following the notations in Section \ref{review}, let ${\C K}={\mathbb X}$ and ${\C F}={\mathbb Y}.$ We construct $\C L$ systematically by first initializing $\mathcal{L}$ by its user-cliques. The user-cliques are indexed by $X\in \mathbb{X}$. For each $X\in \mathbb{X}$ create the vertices corresponding to the user-clique indexed by $X$ as
$C_X \triangleq \left\{(X,Y) :Y\in \mathbb{Y}, \sum\limits_{V_i\in X}V_i +\sum\limits_{V_i\in Y}V_i \in \mathbb{U}\right\}$. Thus, $V({\C L}) \triangleq \bigcup\limits_{X\in{\mathbb X}}C_X.$ Now, for each $Y\in \mathbb{Y}$ we construct the subfile clique of ${\C L}$ associated with $Y$ as $C_{Y}\triangleq \left\{(X,Y) :X\in \mathbb{X}, \sum\limits_{V_i\in X}V_i +\sum\limits_{V_i\in Y}V_i \in \mathbb{U}\right\}$. We thus see that $V({\C L})=\bigcup\limits_{Y\in{\mathbb Y}}C_Y.$ Now, if we show that the user cliques (and equivalently, subfile cliques) are of the same size each, then the properties P1-P3 will be satisfied by $\C L$. By invoking the notations from Section \ref{review}, we have $K=|\mathbb{\mathbb{X}}|$ (number of user-cliques), %size of user-clique $D=|C_V|$, size of subfile-clique $|C_X|$ 
and subpacketization $F=|\mathbb{Y}|$ (the number of subfile cliques).

% \begin{lemma} \label{no of 1D spaces}
% Consider $V \in \mathbb{V}$. Then $|\{T\in \mathbb{T}: T\in V, T\notin W\}|=q^{t-1}$.
% \end{lemma}

% \begin{IEEEproof}
% Since $V$ is a $t$-dim space and $W$ is a $(t-1)$-dim space, we have $|\{T\in \mathbb{T}: T\in V\}|=\theta(t)$ and $|\{T\in \mathbb{T}: T\in W\}|=\theta(t-1)$. Therefore we can write $|\{T\in \mathbb{T}: T\in V, T\notin W\}|=\theta(t)-\theta(t-1)=\frac{q^t-q^{t-1}}{q-1}=q^{t-1}$.
% \end{IEEEproof}

We now find the values of $K,F$, the size of user clique $|C_X|$ and the size of subfile clique $|C_Y|$. 
%%%
\begin{lemma}
\label{K,F,D,c expressions}
\begin{align*}
K =|\mathbb{X}| &= \frac{q}{2}\gbinom{k-t+1}{1} \gbinom{k-t}{1} . \\
F =|\mathbb{Y}| &= \gbinom{k-t+1}{m+1}\frac{\prod\limits_{i=0}^{m}(q^{m+1}-q^{i})}{(m+1)! (q-1)^{(m+1)}}. \\
|C_X| &=\frac{1}{(m+1)!} \hspace{0.2cm}  q^{\frac{(m+1)(m+4)}{2}} \prod\limits_{i=1}^{m+1} \gbinom{k-t-i}{1} . \\
|C_Y| &= \frac{q^{(2m+3)}}{2} \gbinom{k-m-t}{1} \gbinom{k-m-t-1}{1}.
\end{align*}
% \vspace{-0.4cm}
for any $X\in {\mathbb{X}}$, $Y\in {\mathbb{Y}}$.
    %\item[(a4)] $F=\gbinom{k-t+1}{m+1}\dfrac{\prod\limits_{i=0}^{m}(q^{m+1}-q^{i})}{(q-1)^{m+1}} $.

\end{lemma}

\begin{IEEEproof}
(Finding $K=|\mathbb{X}|$): 
Finding $|\mathbb{X}|$ is equivalent to counting the number of distinct sets $ \{T_1,T_2\}$ (such that $ T_i\in \mathbb{T}~ \forall i\in [2]$ and $W\oplus T_1 \oplus T_2 \in  \mathbb{R} $) which gives distinct $\{W\oplus T_1,W\oplus T_2\} \in \mathbb{X}$.
By Lemma \ref{no of sets of LI 1D spaces} we have, the number of distinct sets $ \{T_1,T_2\}$, such that $ T_i\in \mathbb{T}~(\forall i\in [2])$ and  $W\oplus T_1 \oplus T_2 \in  \mathbb{R},$ is 
$\frac{\prod\limits_{i=0}^{1}(\theta(k)-\theta(t-1+i))}{2!}$. It is easy to check that $\{W\oplus T_1,W\oplus T_2\} \in \mathbb{X}$. By Corollary \ref{no of 1D spaces outside a hper space} we have, the number of distinct $T\in \mathbb{T}$ such that $W\oplus T= V$ for some fixed $V\in \mathbb{V}$ is $q^{t-1}$. Therefore for each $\{W\oplus T_1,W\oplus T_2\} \in \mathbb{X}$ there exist $(q^{t-1})^2=q^{2(t-1)}$ distinct $\{T_1',T_2'\}$ (where $T_i'\in \mathbb{T}, \forall i\in [2]$) such that $W\oplus T_i=W\oplus T_i', \forall i\in [2]$.
Therefore we can write 

$K=\frac{\prod\limits_{i=0}^{1}(\theta(k)-\theta(t-1+i))}{2q^{2(t-1)}}= \frac{(q^k-q^{t-1})(q^k-q^t)}{2q^{2(t-1)}(q-1)^2}=\frac{q}{2}.\frac{q^{k-t+1}-1}{q-1}.\frac{q^{k-t}-1}{q-1}=\frac{q}{2}\gbinom{k-t+1}{1}\gbinom{k-t}{1}$.

(Finding $F=|\mathbb{Y}|$): 
Finding $|\mathbb{Y}|$ is equivalent to counting the number of distinct sets $ \{T_1,T_2,\cdots, T_{m+1}\}$ (such that $ T_i\in \mathbb{T}~\forall i\in [m+1]$ and $W\oplus T_1 \oplus T_2 \oplus \cdots \oplus T_{m+1} \in  \mathbb{S} $) which gives distinct $\{W\oplus T_1,W\oplus T_2, \cdots ,W\oplus T_{m+1}\} \in \mathbb{Y}$.
By Lemma \ref{no of sets of LI 1D spaces} we have, the number of distinct sets $ \{T_1,T_2,\cdots,T_{m+1}\}$, such that $ T_i\in \mathbb{T}~(\forall i\in [m+1])$ and $W\oplus T_1 \oplus T_2 \cdots \oplus T_{m+1} \in  \mathbb{S} $, is 
$\frac{\prod\limits_{i=0}^{m}(\theta(k)-\theta(t-1+i))}{(m+1)!}$. It is easy to check that $\{W\oplus T_1,W\oplus T_2, \cdots ,W\oplus T_{m+1}\} \in \mathbb{Y}$. By Corollary \ref{no of 1D spaces outside a hper space} we have, the number of distinct $T\in \mathbb{T}$ such that $W\oplus T= V$ for some fixed $V\in \mathbb{V}$ is $q^{t-1}$. Therefore for each $\{W\oplus T_1,W\oplus T_2, \cdots ,W\oplus T_{m+1}\} \in \mathbb{Y}$ there exist $(q^{t-1})^{m+1}=q^{(m+1)(t-1)}$ distinct $\{T_1',T_2',\cdots, T_{m+1}'\}$ (where $T_i'\in \mathbb{T}~ \forall i\in [m+1]$) such that $W\oplus T_i=W\oplus T_i', \forall i\in [m+1]$. 
Therefore we can write 
\begin{align*}
 F&=\frac{\prod\limits_{i=0}^{m}(\theta(k)-\theta(t-1+i))}{(m+1)! q^{(m+1)(t-1)}}\\
&= \frac{\prod\limits_{i=0}^{m}(q^k-q^{t-1+i})}{(m+1)! q^{(m+1)(t-1)}(q-1)^{m+1}}\\
&= \frac{q^{(m+1)(t-1)}\left(\prod\limits_{i=0}^{m}q^i\right)\left(\prod\limits_{i=0}^{m}(q^{k-t+1-i}-1)\right)}{(m+1)! q^{(m+1)(t-1)}(q-1)^{m+1}} \\
&= \frac{\prod\limits_{i=0}^{m}(q^{k-t+1-i}-1)}{\prod\limits_{i=0}^{m}(q^{m+1-i}-1)} \frac{\left(\prod\limits_{i=0}^{m}(q^{m+1-i}-1)\right)\left(\prod\limits_{i=0}^{m}(q^i)\right)}{(m+1)! (q-1)^{m+1}}\\
&= \gbinom{k-t+1}{m+1} \frac{\prod\limits_{i=0}^{m}(q^{m+1}-q^i)}{(m+1)! (q-1)^{m+1}}.
\end{align*}
%%%

(Finding $|C_X|$): 
Consider an arbitrary $X=\{V_a,V_b\}\in \mathbb{X}$. We have $V_a+V_b=R$, for some $R \in \mathbb{R}$. We know that $dim(R)=t+1$. 
Now, finding $|C_X|$ is equivalent to counting the number of distinct sets $ \{T_1,T_2,\cdots, T_{m+1}\}$ (such that $ T_i\in \mathbb{T}, \forall i\in [m+1], R\oplus T_1 \oplus T_2 \oplus \cdots \oplus T_{m+1} \in  \mathbb{U} $) which gives distinct $\{W\oplus T_1,W\oplus T_2, \cdots ,W\oplus T_{m+1}\} \in \mathbb{Y}$.
By Lemma \ref{no of sets of LI 1D spaces} we have, the number of distinct sets $ \{T_1,T_2,\cdots,T_{m+1}\}$ such that $ T_i\in \mathbb{T}, \forall i\in [m+1], R\oplus T_1 \oplus T_2 \oplus \cdots \oplus T_{m+1} \in  \mathbb{U} $ is 
$\frac{\prod\limits_{i=0}^{m}(\theta(k)-\theta(t+1+i))}{(m+1)!}$. It is easy to check that $\{W\oplus T_1,W\oplus T_2, \cdots, W\oplus T_{m+1}\} \in \mathbb{Y}$. By Corollary \ref{no of 1D spaces outside a hper space} we have, the number of distinct $T\in \mathbb{T}$ such that $W\oplus T= V$ for some fixed $V\in \mathbb{V}$ is $q^{t-1}$. Therefore for each $\{W\oplus T_1,W\oplus T_2, \cdots, W\oplus T_{m+1}\} \in \mathbb{Y}$ there exist $(q^{t-1})^{m+1}=q^{(m+1)(t-1)}$ distinct $\{T_1',T_2',\cdots, T_{m+1}'\}$ (where $T_i'\in \mathbb{T}, \forall i\in [m+1]$) such that $W\oplus T_i=W\oplus T_i', \forall i\in [m+1]$.
Therefore we can write
\begin{align*}
|C_X|&=\frac{\prod\limits_{i=0}^{m}(\theta(k)-\theta(t+1+i))}{(m+1)! q^{(m+1)(t-1)}}\\
&= \frac{\prod\limits_{i=0}^{m}(q^k-q^{t+1+i})}{(m+1)! q^{(m+1)(t-1)}(q-1)^{m+1}}
\end{align*}
\begin{align*}
&= \frac{q^{(m+1)(t+1)}\left(\prod\limits_{i=0}^{m}q^i\right)\left(\prod\limits_{i=0}^{m}(q^{k-t-1-i}-1)\right)}{(m+1)! q^{(m+1)(t-1)}(q-1)^{m+1}} \\
&=  \frac{q^{2(m+1)}q^{\frac{m(m+1)}{2}}}{(m+1)! } \frac{\prod\limits_{i=1}^{m+1}(q^{k-t-i}-1)}{(q-1)^{m+1}} \\
&= \frac{1}{(m+1)!}q^{\frac{(m+1)(m+4)}{2}} \prod\limits_{i=1}^{m+1}\gbinom{k-t-i}{1}.
\end{align*}

(Finding $|C_Y|$): Consider an arbitrary $Y=\{V_1,V_2,\cdots,V_{m+1}\}\in \mathbb{Y}$. We have $\sum\limits_{i=1}^{m+1}V_i=S$, for some $S \in \mathbb{S}$. We know that $dim(S)=t+m$.
Now, finding $|C_Y|$ is equivalent to counting the number of distinct sets $ \{T_1,T_2\}$ (such that $ T_i\in \mathbb{T}, \forall i\in [2], S\oplus T_1 \oplus T_2 \in  \mathbb{U} $) which gives distinct $\{W\oplus T_1,W\oplus T_2\} \in \mathbb{X}$.
By Lemma \ref{no of sets of LI 1D spaces} we have, the number of distinct sets $ \{T_1,T_2\}$ such that $ T_i\in \mathbb{T}, \forall i\in [2], S\oplus T_1 \oplus T_2 \in  \mathbb{U} $ is 
$\frac{\prod\limits_{i=0}^{1}(\theta(k)-\theta(t+m+i))}{2!}$. It is easy to check that $\{W\oplus T_1,W\oplus T_2\} \in \mathbb{X}$. By Corollary \ref{no of 1D spaces outside a hper space} we have, the number of distinct $T\in \mathbb{T}$ such that $W\oplus T= V$ for some fixed $V\in \mathbb{V}$ is $q^{t-1}$. Therefore for each $\{W\oplus T_1,W\oplus T_2\} \in \mathbb{X}$ there exist $(q^{t-1})^2=q^{2(t-1)}$ distinct $\{T_1',T_2'\}$ (where $T_i'\in \mathbb{T}, \forall i\in [2]$) such that $W\oplus T_i=W\oplus T_i', \forall i\in [2]$.
Therefore we can write 
\begin{align*}
    |C_Y|&=\frac{\prod\limits_{i=0}^{1}(\theta(k)-\theta(t+m+i))}{2 q^{2(t-1)}}= \frac{\prod\limits_{i=0}^{1}(q^k-q^{t+m+i})}{2 q^{2(t-1)}(q-1)^{2}}\\
    &= \frac{q^{t+m}(q^{k-t-m}-1)q^{t+m+1}(q^{k-t-m-1}-1)}{2 q^{2(t-1)}(q-1)^{2}} \\
    &=  \frac{q^{2m+3}}{2}\gbinom{k-m-t}{1}\gbinom{k-m-t-1}{1}.
\end{align*}

This completes the proof.
\end{IEEEproof}
%%%

\begin{remark}
It can be checked that $K|C_X|=F|C_Y|$ $ \quad \qquad (=|V({\C L})|$ by construction$).$
\end{remark}

Note that by Lemma \ref{K,F,D,c expressions}, we have the size of the subfile cliques of ${\cal L}$ as $|C_Y|$ (for any $Y\in{\mathbb Y}$), and this is same for each $Y$. Similarly the user-cliques all have the same size $|C_X|$. Hence $\C L$ satisfies properties P1-P3. We now show that $\Lbarsq$ has a clique cover with $d$-sized disjoint cliques for some $d$. Therefore ${\cal L}$ is in fact a $(c=|C_Y|,d)$-caching line graph, giving rise to the main result in this section which is Theorem \ref{main result}. 
\vspace{-0.2cm}
\subsection{Delivery Scheme from a clique cover of $\Lbarsq$}
We first describe a clique of $\Lbarsq$ and show that such equal-sized cliques partition $V(\mathcal{L})=V(\Lbarsq)$. This will suffice to show the delivery scheme as per Theorem $2$ of \cite{PK} (summarized in Remark \ref{cdremark}  in Section \ref{review} in this work). 

We now present a clique of size $\binom{m+3}{2}$ in ${\Lbarsq}$ (where $\binom{a}{b}$ represents binomial coefficient). Recall the definition of $\mathbb Z$ from (\ref{eqn Z}). 
%%%
\begin{lemma} 
\label{size of transmission clique}
Consider $Z=\{V_1,V_2,\cdots,V_{m+3}\}\in \mathbb{Z}$. Then
$C_Z=\left\{(\{V_i,V_j\},Z\setminus\{ V_i,V_j\}),\forall V_i,V_j\in Z,i\neq j\right\}\subseteq V(\Lbarsq)$ is a clique in $\overline{\mathcal{L}^2}$.
\end{lemma}
%%%
\begin{IEEEproof}
First note that $C_Z$ is well defined as $\sum\limits_{i=1}^{m+3}V_i \in \mathbb{U}$  and hence $(\{V_i,V_j\},Z\setminus \{V_i,V_j\})\in V({\C L})$ (belongs to user clique $C_{\{V_i,V_j\}}$). To show that the set of vertices of $\C L$ in $C_Z$ forms a clique of $\Lbarsq$, we have to show that between any two vertices of $C_Z$, there is an edge in $\Lbarsq$. For this purpose, we use Lemma \ref{edge in compliment of square}. Consider two distinct vertices $(\{V_i,V_j\},Z\setminus \{V_i,V_j\}), (\{V_{i'},V_{j'}\},Z\setminus \{V_{i'},V_{j'}\})\in C_Z$. Without loss of generality, let $V_i\notin \{V_{i'}.V_{j'}\}$. Then it is clear that $\{V_i,V_j\}\cap (Z\setminus \{V_{i'},V_{j'}\})$ contains $V_i$. Thus we have $V_i+V_j+\sum\limits_{V_l\in Z\setminus \{V_{i'},V_{j'}\}}V_l \notin \mathbb{U}$ (as $\mathbb U$ contains only $m+t+2$ dimensional subspaces, however $V_i+V_j+\sum\limits_{V_l\in Z\setminus \{V_{i'},V_{j'}\}}V_l$ has dimension at most $m+t+1$ as $V_i$ appears twice in this sum). Similarly we can show that $V_{i'}+V_{j'}+\sum\limits_{V_l\in Z\setminus \{V_i,V_j\}}V_l \notin \mathbb{U}$. Therefore we have that the ordered pairs $(\{V_i,V_j\},Z\setminus \{V_{i'},V_{j'}\}), (\{V_{i'},V_{j'}\},Z\setminus \{V_i,V_j\})$ are not present in $V(\overline{\mathcal{L}^2})$. By invoking Lemma \ref{edge in compliment of square}, $\left\{(\{V_i,V_j\},Z\setminus \{V_i,V_j\}), (\{V_{i'},V_{j'}\},Z\setminus \{V_{i'},V_{j'}\})\right\} \in E(\Lbarsq)$. As we started from arbitrary vertices in $C_Z$ and showed that there is an edge of $\Lbarsq$ containing both, this proves that $C_Z$ forms a clique in $\Lbarsq$. It is easy to see that $|C_Z|=\binom{m+3}{2}$. Hence proved.
\end{IEEEproof}

Now we show that the cliques $\{C_Z : Z\in \mathbb{Z}\}$ partition $V({\Lbarsq})$.
\begin{lemma} \label{partition of L}
$\bigcup\limits_{Z\in \mathbb{Z}}C_Z=V(\mathcal{L})=V(\Lbarsq)$,
where this union is a disjoint union (the cliques $C_Z$ are as defined in Lemma \ref{size of transmission clique}).
\end{lemma}
\begin{IEEEproof}
Consider $Z, Z'\in \mathbb{Z}$ such that $Z\neq Z'$. By definition of $C_Z,C_{Z'}$, we have $C_Z \cap C_{Z'}=\phi$. Now consider an arbitrary vertex $(\{V_1,V_2\},\{V_3,\cdots,V_{m+3}\}) \in V(\mathcal{
L})$. By the construction of $\mathcal{L}$,  $ \sum\limits_{i=1}^{m+3}V_i \in \mathbb{U}$. Therefore $(\{V_1,V_2\},\{V_3,\cdots,V_{m+3}\})$ lies in the unique clique, $ C_{\{V_1,V_2,\cdots,V_{m+3}\}} $ (defined as in Lemma \ref{size of transmission clique}). Hence proved.
\end{IEEEproof}

Finally we present our coded caching scheme using the caching line graph constructed above.

\begin{theorem}\label{main result}
The caching line graph $\mathcal{L}$ constructed above is a $ \left(c=\frac{q^{(2m+3)}}{2} \gbinom{k-m-t}{1} \gbinom{k-m-t-1}{1}, d=\binom{m+3}{2} \right)$-caching line graph and defines a coded caching scheme with 

\[K=\frac{q}{2}\gbinom{k-t+1}{1} \gbinom{k-t}{1},\] 
\[F=\gbinom{k-t+1}{m+1}\dfrac{\prod\limits_{i=0}^{m}(q^{m+1}-q^{i})}{(m+1)! (q-1)^{m+1}},\] 

\[\frac{M}{N}=1-q^{2(m+1)}\dfrac{\gbinom{k-m-t}{1}\gbinom{k-m-t-1}{1}}{{\gbinom{k-t+1}{1}\gbinom{k-t}{1}}},\] 
 
\[R=\dfrac{q^{(2m+3)}}{(m+2)(m+3)}\gbinom{k-m-t}{1}\gbinom{k-m-t-1}{1}.\]
\end{theorem}
%%%
\begin{IEEEproof}
From Lemma \ref{K,F,D,c expressions}, we get the expression of $K$ and $F.$
%$K=\gbinom{k-t+1}{1}$, size of subfile clique  $c=q^{(m+1)}\gbinom{k-m-t}{1}$, size of user clique $D=\gbinom{k-t}{m+1}\dfrac{q^{m+1}\prod\limits_{i=0}^{m}(q^{m+1}-q^{i})}{(q-1)^{m+1}}$. 
Further we see that the subfile cliques partition the vertices of ${\cal L}$ by definition and also $c=|C_Y|$ for any $Y\in \mathbb{Y}$ (the size of each subfile clique). By Lemma
\ref{size of transmission clique} and Lemma \ref{partition of L}, the size of the cliques of  $\Lbarsq$ is $\binom{m+3}{2}$ and they partition the vertices. Hence  ${\C L}$ is a  $ \left(c=\frac{q^{(2m+3)}}{2} \gbinom{k-m-t}{1} \gbinom{k-m-t-1}{1}, d=\binom{m+3}{2} \right)$-caching line graph. 
%Using the fact that $F=\frac{KD}{c}$ by Theorem \ref{cliquecoverlinegraph}, and since $\scriptsize \frac{\gbinom{k-t+1}{1}}{\gbinom{k-m-t}{1}}= \frac{\gbinom{k-t+1}{m+1}}{\gbinom{k-t}{m+1}}$, we obtain the expression for $F$ in this theorem. The other parameters similarly follow by Theorem \ref{cliquecoverlinegraph}.

Thus, we have by Theorem 2 of \cite{PK
} (paraphrased in Remark \ref{cdremark} in Section \ref{review} in this work),

\[\frac{M}{N}=1-\frac{c}{K}=1-q^{2(m+1)}\dfrac{\gbinom{k-m-t}{1}\gbinom{k-m-t-1}{1}}{{\gbinom{k-t+1}{1}\gbinom{k-t}{1}}}.\]

\[R=\frac{c}{d}=\dfrac{q^{(2m+3)}}{(m+2)(m+3)}\gbinom{k-m-t}{1}\gbinom{k-m-t-1}{1}.\]
This completes the proof.
\end{IEEEproof}

%%%%%%%%% ALGORITHM

We now present Algorithm \ref{algorithm}, which presents the caching and delivery scheme developed in this section. For the given $K' ,M' ,N'$ select the appropriate  parameters $k,m,t,q$ which give $K,\frac{M}{N},F,R$ such that $(K- K')$ and $(\frac{M'}{N'}-\frac{M}{N})$ are non negative and as small as possible (we treat the extra users $K-K'$ as dummy users). Now construct a $(c,d)$-caching line graph $(\mathcal{L})$ as mentioned in Section \ref{our scheme B subsection} and find $\mathbb{X}$ (user indices), $\mathbb{Y}$ (subfile indices), $\mathbb{Z}$ (indices of cliques of $\Lbarsq$, equivalently indices of transmissions$)$ by using (\ref{eqn X}),(\ref{eqn Y}),(\ref{eqn Z}).

\begin{algorithm}
\caption{Coded caching scheme proposed in Theorem \ref{main result}}
\label{algorithm}
\begin{algorithmic}[1]
\Procedure{Placement Phase}{}
    \For{each $i\in [N']$}
    \State Split $W_i$ into $\{W_{i,Y}:Y\in\mathbb{Y}\}$.
    \EndFor
    \For {each $X \in \mathbb{X}$}
    \State user $X$ caches the subfiles $W_{i,Y}, \forall i\in [N'], \forall Y\in \mathbb{Y}$ such that $(X,Y)\notin V(\mathcal{L})$.
    \EndFor
\EndProcedure
\Procedure{Delivery Phase}{ demand of user $X$ is represented as $W_{d_{X}}, \forall X\in \mathbb{X}$}
    \For {each $ Z=\{V_1,V_2,\cdots,V_{m+3}\} \in \mathbb{Z}$ }
    \State Server transmits $\sum\limits_{\{V_i,V_j\}\subset Z} W_{d_{\{V_i,V_j\}},Z\setminus \{V_i,V_j\}}$.
    \EndFor
\EndProcedure
\end{algorithmic}
\end{algorithm}

\section{Asymptotic analysis of the proposed scheme}
\label{asymptotics}

%TABLE WITH ALL VALUES OF K,U
% Table for coded caching schemes comparison

\newcolumntype{g}{>{\columncolor{Gray}}c}
\begin{table*}
\vspace{0.2cm}
\centering
\begin{tabular}{|g|c|c||g|c|c||g|c|c||g|c|c|}
\hline
%$(k,m,t,q)$ & $(m',q')$ & & & & & & \\ 
$K_1$ & $ K_2$ & $K_3$ & $U_1$  & $U_2$ & $U_3$ &$F_1$ &  $F_2$ & $F_3$ & $\gamma_1$& $\gamma_2$ & $\gamma_3$\\
%\scriptisize [this work] & \cite{YCTCPDA} & \scriptisize[this work] & \cite{YCTCPDA}& \scriptisize[this work]& \cite{YCTCPDA}&\scriptisize[this work]& \cite{YCTCPDA}\\\
%$  $ & $q'(m'+1)$ & %$ \frac{q^{m+1}\gbinom{k-t}{m+1}}{\gbinom{k-t+1}{m+1}}$ 
%& $1-\frac{1}{q'}$ & %$ \gbinom{k-t+1}{m+1}\dfrac{\prod\limits_{i=0}^{m}(q^{m+1}-q^{i})}{(q-1)^{m+1}}$ 
%& $q^{m'}$ & $(m+2)$ & $(\frac{MK}{N})$\\
&\cite{haribhavanaprasad}&\cite{YCTCPDA} & &\cite{haribhavanaprasad}& \cite{YCTCPDA} & & \cite{haribhavanaprasad}& \cite{YCTCPDA} & &\cite{haribhavanaprasad}&\cite{YCTCPDA}\\
%\scriptisize [proposed] & \cite{YCTCPDA} & \scriptisize[proposed] & \cite{YCTCPDA}& \scriptisize[proposed]& \cite{YCTCPDA}&\scriptisize[proposed]& \cite{YCTCPDA}\\
\hline
%$(10,2,2,2)$ & $(6,73)$ &&&&&&\\
8001 & 8191 & 8008 & 0.93 & 0.94 & 0.93 & 8001 & $10^{29}$ & inf & 6 & 10 & 572
\\
\hline

8001 & 8191 & 8001 & 0.67 & 0.75 & 0.67 & $10^7$ & $10^{35}$ & inf & 15 & 12 & $10^3$
\\
\hline
%$(9,3,2,2)$ & $(14,17)$ &&&&&&\\
780 & 781 & 780 & 0.62 & 0.80 &  0.67 & 780 & $10^{10}$ & $10^{123}$ & 6 & 5 & 260
\\
\hline

%$(8,3,2,2)$ & $(13,9)$ &&&&&&\\
465 & 511 & 468 & 0.72 & 0.75 & 0.75 & 465 & $10^{15}$ & $10^{69}$ & 6 & 8 & 117
\\
\hline

%$(9,4,3,2)$ & $(31,4)$ &&&&&&\\
105 & 127 & 104 & 0.46 & 0.50 & 0.50 & 105 & $10^{9}$ & $10^{15}$ & 6 & 7 & 52
\\
\hline

% %$(7,3,2,2)$ &$(15,4)$ &&&&&&\\
% 63 & 64& 0.76 & 0.75 & $ 10^7$ & $10^9 $ & 5 &16\\
% \hline

% %$(7,3,3,3)$ & (39,3) &&&&&&\\
% 121 & 120 & 0.67 & 0.66 & $10^8$ & $10^{18} $ & 5 & 40\\
% \hline
% %$(6,3,2,2)$ & $(14,2)$ &&&&&&\\
% 31 & 30 & 0.51 & 0.50 & $10^5$ & $10^4 $ & 5& 15\\
% \hline
\end{tabular}
\caption{Comparison of Coded caching schemes presented in \cite{haribhavanaprasad},\cite{YCTCPDA} with this work. (inf represents $> 10^{307}$).
%For some specific values of $K,U=1-\frac{M}{N}$, we compare the results of \cite{haribhavanaprasad}, \cite{YCTCPDA} with this work.
}
%Here $\alpha$ represents our scheme and $\beta$ represents scheme proposed in \cite{YCTCPDA}.
%$K,F,U,G$ with subscript $\alpha$ denote the parameters for the scheme presented in this paper whereas parameters with subscript $\beta$ characterize the performance of PDA scheme CITE HERE.
\label{tab1}
\end{table*}

%%%%%%%%%%%%%%%%%%%%%

% Table for D2D coded caching schemes comparison

\begin{table*}
\centering
\begin{tabular}{|g|c|c||g|c|c||g|c|c||g|c|c|}

\hline

$K_1^\mathcal{D}$ & $K_2^\mathcal{D}$ & $K_3^\mathcal{D}$ & $U_1^\mathcal{D}$ & $U_2^\mathcal{D}$ & $U_3^\mathcal{D}$ &
$F_1^\mathcal{D}$ & $F_2^\mathcal{D}$ & $F_3^\mathcal{D}$ & $R_1^\mathcal{D}$ & $R_2^\mathcal{D}$ & $R_3^\mathcal{D}$

\\

&\cite{hypercubeD2D}&\cite{D2D} & &\cite{hypercubeD2D}&\cite{D2D} & 
&\cite{hypercubeD2D}&\cite{D2D} & &\cite{hypercubeD2D}&\cite{D2D}

\\

\hline

8001 & 8001 & 8001 & 0.93 & 0.98 & 0.93 & 40005 & $10^{174}$ & inf & 1488 & 89.44 & 13.28
\\

\hline

7260 & 7260 & 7260 & 0.87 & 0.98 & 0.87 & 36300 & $10^{164}$ & inf & 1263 & 85.20 & 6.70
\\

\hline

1953 & 1953 & 1953 & 0.86 & 0.97 & 0.86 & 9765 & $10^{72}$ & inf & 336 & 44.19 & 6.15
\\

\hline

780 & 780 & 780 & 0.62 & 0.96 & 0.62 & 3900 & $10^{40}$ & $10^{225}$ & 97.2 & 27.92 & 1.65
\\

\hline

465 & 465 & 465 & 0.72 & 0.95 & 0.72 & 2325 & $10^{28}$ & $10^{119}$ & 67.2 & 21.56 & 2.60
\\

\hline

105 & 105 & 105 & 0.46 & 0.90 & 0.46 & 525 & $10^{10}$ & $10^{32}$ & 9.6 & 10.25 & 0.84
\\

\hline

\end{tabular}
\caption{Comparison of D2D Coded caching schemes presented in \cite{hypercubeD2D},\cite{D2D} with this work. (inf represents $> 10^{307}$).
%For some specific values of $K,U=1-\frac{M}{N}$, we compare the results of \cite{haribhavanaprasad}, \cite{YCTCPDA} with this work.
}

\label{tab2}
\end{table*}

%%%%%%%%%%%%%%%%%%%%%%%%%

In this section, we analyse the asymptotic behaviour of $F,R$ for our coded caching scheme proposed in Theorem \ref{main result} as $\frac{M}{N}$ is upper bounded by a constant and $K\rightarrow \infty$. We show that $F=q^{O((log_qK)^2)}$, while $R=\Theta(\frac{K}{(log_qK)^2})$. Towards this end, we first recall some bounds on  $q$-binomial coefficients. 
% A proof is available in \cite{PK}.
%%%
\begin{lemma}\cite{PK}
\label{approximations}
Let $a,b,f \in \mathbb{Z}^+$ and $q$ be some prime power. Then,  
% $\small q^{(a-b)b}\leq \gbinom{a}{b} \leq q^{(a-b+1)b}$ and $\small q^{(a-f-b-1)\delta} \leq \frac{\gbinom{a}{b}}{\gbinom{a}{f}} \leq  q^{(a-f-b+1)\delta},$ where $\scriptsize \delta=|b-f|$. 
% \begin{align*}
% \label{eqn31}
$q^{(a-b)b}\leq \gbinom{a}{b} \leq  q^{(a-b+1)b}$.
%\\
% \label{eqn33}
% q^{(a-f-b-1)\delta} &\leq \frac{\gbinom{a}{b}}{\gbinom{a}{f}} \leq  q^{(a-f-b+1)\delta}, \\
% % \nonumber
% \textit{where } \delta=|b-f|.
% \end{align*}
% 

% \label{eqn32}
% &q^{(a-f-1)b} &\leq &\frac{\gbinom{a}{b}}{\gbinom{f}{b}}\leq & q^{(a-f+1)b}\\

\end{lemma}

Throughout our analysis we assume $q$ is constant. We now upper bound $\frac{M}{N}$ by a constant.
% proceed to analyse the asymptotics of the scheme. 
We have by Theorem \ref{main result},
\begin{align*}
\label{eqn34}
1-\frac{M}{N}&=1-q^{2(m+1)}\dfrac{\gbinom{k-m-t}{1}\gbinom{k-m-t-1}{1}}{{\gbinom{k-t+1}{1}\gbinom{k-t}{1}}} \\ &=q^{2(m+1)}\frac{(q^{k-m-t}-1)(q^{k-m-t-1}-1)}{(q^{k-t+1}-1)(q^{k-t}-1)} \\
&\geq \frac{(q^{k-t+1}-q^{m+1})(q^{k-t}-q^{m+1})}{(q^{k-t+1})(q^{k-t})} \\
&= \left(1-\frac{1}{q^{k-t-m}}\right)\left(1-\frac{1}{q^{k-t-m-1}}\right).
\end{align*}
To lower bound $1-\dfrac{M}{N}$ by a constant, let $k-m-t=
\alpha$, where $\alpha$ is a constant. Note that $\alpha \geq 2$ as $k \geq m+t+2$.
Thus we have,
\begin{align*}
    1-\frac{M}{N} &\geq \left(1-\frac{1}{q^\alpha}\right)\left(1-\frac{1}{q^{\alpha-1}}\right)%\geq 1-\frac{2}{q^{\alpha-1}} %\\
    %&= 1-\frac{1}{q^{\alpha-1}}-\frac{1}{q^{\alpha}}+\frac{1}{q^{2\alpha-1}} \\
    %&\geq 1-\frac{1}{q^{\alpha-1}}-\frac{1}{q^{\alpha}} = 1-\frac{1}{q^{\alpha}}(q+1) \\
    %&\geq 1-\frac{1}{q^{\alpha}}(q+q) =1-\frac{2}{q^{\alpha-1}}
\end{align*}
Therefore $\frac{M}{N}\leq \frac{2}{q^{\alpha -1}}$.

We have $K=\frac{q}{2}\gbinom{k-t+1}{1} \gbinom{k-t}{1}$. We analyse our scheme as $(k-t)$ grows large (thus $K$ grows large). By Lemma \ref{approximations}, we have 
\begin{align*}
\frac{q}{2}.q^{k-t}.q^{k-t-1}&\leq K \leq \frac{q}{2}.q^{k-t+1}.q^{k-t},  \\
q^{2(k-t)}&\leq 2K\leq q^{2(k-t+1)},  \\
2(k-t) &\leq \log_q{2K} \leq 2(k-t+1).
\end{align*}

Hence we have
\begin{equation}
\label{eqn k-t bounds}
    \frac{1}{2}\log_q{2K}-1 \leq k-t \leq \frac{1}{2}\log_q{2K},
\end{equation}
\begin{equation}
\label{eqn (k-t)^2 bounds}
    \frac{1}{4}(\log_q{2K})^2-\log_q{2K}+1 \leq (k-t)^2 \leq \frac{1}{4}(\log_q{2K})^2.
\end{equation}
We now get the asymptotics for the rate. The rate expression in Theorem \ref{main result} can be written as 
$R=\dfrac{K(1-\frac{M}{N})}{d} = \dfrac{2K(1-\frac{M}{N})}{(m+2)(m+3)}$.

Now we have,
$(m+2)(m+3)=m^2+5m+6=(k-t-\alpha)^2+5(k-t-\alpha)+6=(k-t)^2+(5-2\alpha)(k-t)+(\alpha^2-5\alpha+6)$.

Therefore by using (\ref{eqn k-t bounds}) and (\ref{eqn (k-t)^2 bounds}) we have
$\frac{2K(1-\frac{M}{N})}{\frac{1}{4}(\log_q{2K})^2+\frac{5-2\alpha}{2}\log_q{2K}+\alpha^2-5\alpha+6} \leq R \leq \frac{2K(1-\frac{M}{N})}{\frac{1}{4}(\log_q{2K})^2+\frac{3-2\alpha}{2}\log_q{2K}+\alpha^2-3\alpha+2}.$

After some simple manipulations, we see that $\qquad \qquad R = \Theta(\frac{K}{(\log_q{2K})^2})=\Theta(\frac{K}{(\log_q{K})^2})$.

We now obtain the asymptotics for subpacketization $F$. 
From initial expressions for $F,K$ in proof of Lemma \ref{K,F,D,c expressions}, we have $F=\frac{\prod\limits_{i=0}^{m}(\theta(k)-\theta(t-1+i))}{(m+1)! q^{(m+1)(t-1)}}$ and $K=\frac{\prod\limits_{i=0}^{1}(\theta(k)-\theta(t-1+i))}{2q^{2(t-1)}}$.
Therefore ,
\begin{align*}
\frac{F}{K} &=\frac{2}{(m+1)! q^{(m-1)(t-1)}}\prod\limits_{i=2}^{m}(\theta(k)-\theta(t-1+i))  \\
&= \frac{2}{(m+1)! q^{(m-1)(t-1)}}\prod\limits_{i=1}^{m-1}(\theta(k)-\theta(t+i))  \\
&= \frac{2}{(m+1)! q^{(m-1)(t-1)}}\prod\limits_{i=1}^{m-1}\left(\frac{q^k-q^{t+i}}{q-1}\right)  \\
&= \frac{2q^{t(m-1)}(\prod\limits_{i=1}^{m-1}q^i)}{(m+1)! q^{(m-1)(t-1)}}\prod\limits_{i=1}^{m-1}\left(\frac{q^{k-t-i}-1}{q-1}\right)  \\
&= \frac{2q^{m-1}(\prod\limits_{i=1}^{m-1}q^i)}{(m+1)! }\prod\limits_{i=1}^{m-1}\gbinom{k-t-i}{1}.
\end{align*}
By Lemma \ref{approximations} we have,
\begin{align*}
\frac{F}{K} &\leq \frac{2q^{m-1}(\prod\limits_{i=1}^{m-1}q^i)}{(m+1)! }\prod\limits_{i=1}^{m-1}q^{k-t-i} \end{align*}

\[\frac{F}{K} \leq \frac{2q^{m-1}}{(m+1)! }\prod\limits_{i=1}^{m-1}q^{k-t-i+i} = \frac{2q^{m-1}q^{(k-t)(m-1)}}{(m+1)!}\] 
Hence we have,
\begin{align*}
\frac{F}{K} &\leq \frac{2}{(m+1)!}q^{(k-t+1)(m-1)}
\end{align*}

% Consider, $(k-t+1)(m-1)=(k-t+1)(k-t-\alpha-1)$.
Since, $m=k-t-\alpha$ and by $(\ref{eqn k-t bounds}),(\ref{eqn (k-t)^2 bounds})$ we have 
% \[\frac{1}{2}\log_q{2K}-1 \leq k-t \leq \frac{1}{2}\log_q{2K}.\]
% \[\frac{1}{2}\log_q{2K} \leq k-t+1 \leq \frac{1}{2}\log_q{2K}+1.\]
% \[\frac{1}{2}\log_q{2K}-2 -\alpha \leq k-t-\alpha-1 \leq \frac{1}{2}\log_q{2K}-\alpha-1.\]

% From this we can write,
% $\frac{1}{4}(\log_q{2K})^2-\frac{(2+\alpha)}{2}\log_q{2K} \leq (k-t+1)(m-1) \leq \frac{1}{4}(\log_q{2K})^2-\frac{\alpha}{2}\log_q{2K}-\alpha-1$.

% $q^{\left(\frac{1}{4}(\log_q{2K})^2-\frac{(2+\alpha)}{2}\log_q{2K}\right)}
$q^{(k-t+1)(m-1)} \leq q^{\left(\frac{1}{4}(\log_q{2K})^2-\frac{\alpha}{2}\log_q{2K}-\alpha-1\right)}$.

% consider, $m+1=k-t-\alpha+1$. Thus we have 
% \[\frac{1}{2}\log_q{2K}-1 \leq k-t \leq \frac{1}{2}\log_q{2K}.\]
% \[\frac{1}{2}\log_q{2K}-\alpha \leq m+1 \leq \frac{1}{2}\log_q{2K}-\alpha+1.\]
% \[\left\lfloor\frac{1}{2}\log_q{2K}-\alpha \right\rfloor ! \leq (m+1)! \leq \left\lceil\frac{1}{2}\log_q{2K}-\alpha+1\right\rceil !\]
% \[\frac{1}{\left\lceil\frac{1}{2}\log_q{2K}-\alpha+1\right\rceil !}\leq
Also, $\frac{1}{(m+1)!} = \frac{1}{(k-t-\alpha+1)!} \stackrel{(\ref{eqn k-t bounds})}{\leq} \frac{1}{\left\lfloor\frac{1}{2}\log_q{2K}-\alpha \right\rfloor !}$.

Therefore,
\[\frac{F}{K}\leq \frac{2q^{\left(\frac{1}{4}(\log_q{2K})^2-\frac{\alpha}{2}\log_q{2K}-\alpha-1\right)}}{\left\lfloor\frac{1}{2}\log_q{2K}-\alpha \right\rfloor !}\]
\[F\leq \frac{q^{\log_q{2K}}q^{\left(\frac{1}{4}(\log_q{2K})^2-\frac{\alpha}{2}\log_q{2K}-\alpha-1\right)}}{\left\lfloor\frac{1}{2}\log_q{2K}-\alpha \right\rfloor !}.\]
Using Stirling's approximation for $x!$ as $\sqrt{2\pi x}\left(\frac{x}{e}\right)^x$ for large $x$, and after some simple manipulations we see that $F=q^{O((\log_q{K})^2)}$.

Finally in Table \ref{tab1}, we compare numerically the scheme in Theorem \ref{main result} with the scheme in \cite{haribhavanaprasad} and \cite{YCTCPDA} for some choices of $K,U=1-\frac{M}{N},F$ and $\gamma$ (the global caching gain, i.e., $\frac{K(1-\frac{M}{N})}{R}$, where $R$ is the rate achieved by the scheme). 
% Note that the constructions in this work and in those of \cite{haribhavanaprasad,YCTCPDA} are parametrized by different quantities. Due to space considerations, we omit them and hence the values of $K$ and $U$ for all the three are chosen to be approximately close. Also, many of the values of the subpacketization $F$ indicate only the exponent obtained, unless the value of $F$ itself is quite small.

We label the parameters of our scheme in Theorem \ref{main result} as $K_1,U_1,F_1,\gamma_1$ where $\gamma_1=d$. The parameters of the scheme presented in \cite{haribhavanaprasad} are labeled as $K_2,U_2,F_2,\gamma_2$( for explicit expressions, the reader is referred to \cite{haribhavanaprasad}). Parameters of the scheme presented in \cite{YCTCPDA} are $K_3=q'(m'+1),U_3=1-\frac{1}{q'},F_3=(q')^{(m')},\gamma_3=\frac{K(1-\frac{M}{N})}{q'-1}$ where $q'(\geq 2),m'\in \mathbb{Z}^+$. As subpacketization can be very large, we approximate it to the nearest positive power of $10$.

% Due to space constraint we only specify $k,m,t,q$ and $q',m'$ from which we are obtaining the parameters in the first row of the Table. For $(k=7,m=1,t=1,q=2)$ the parameters $K_1,U_1,F_1,\gamma_1$ are obtained. $(k=24,m=8,t=12,q=2)$ the parameters $K_2,U_2,F_2,\gamma_2$ are obtained. For $(q'=14,m'=571)$ the parameters $K_3,U_3,F_3,\gamma_3$ are obtained. 

% The first column lists $(k,m,t,q),K_1$ according to the Theorem \ref{main result}. The second column lists $(m',q'),K_2$ parameters of the scheme in \cite{YCTCPDA}. 
%Since we can not match the parameters from this work and \cite{haribhavanaprasad},\cite{YCTCPDA} exactly, we choose approximately equal values. 

We see from the table that the proposed scheme performs much better than \cite{haribhavanaprasad},\cite{YCTCPDA} in terms of the subpacketization (note that the subpacketization of \cite{YCTCPDA} is less than that of the original scheme, \cite{MaN}).
% but pays a price in terms of the rate. 
In particular, the subpacketization obtained by our scheme is much lower than that of \cite{haribhavanaprasad} (which is much lesser than \cite{YCTCPDA}), even for thousands of clients it remains reasonable and practical. The global caching gain however is close to that of \cite{haribhavanaprasad}, and few orders of magnitude smaller than that in \cite{YCTCPDA} (thus the rate of our scheme is comparable to \cite{haribhavanaprasad} and larger than \cite{YCTCPDA}. This indicates that our scheme can be implemented in the practical broadcast coded
caching networks.
%\vspace{-0.3cm}
% \section{Conclusion}
% \label{conclusion}
% In this paper we have presented an explicit construction of coded caching scheme with rate $\Theta \left(  \frac{K}{\log_q{K}}\right)$ and subpacketization $q^{O(\log_{q}{K})^2}$ when uncached fraction is lower bounded a constant. We have also presented a subpacketization dependent lower bound for small caches.
% More schemes can be developed by leveraging the properties of projective geometry and subspace designs, which is an interesting direction for further research.
%%
%%
%    1-\frac{M}{N}&=\frac{q^{(m+1)}(q^{(k-t-m)}-1)}{(q^{(k-t+1)}-1)} \leq \frac{q^{(m+1)}q^{(k-t-m)}}{K(q-1)} \\
%    &\leq \dfrac{q^{(k-t+1)}}{K}= \frac{c_2 (m+2)}{K}  \leq \frac{c_2(\log_q{K}+2)}{K}
%\end{align*}
%Similarly  $1-\frac{M}{N}\geq \frac{c_3\log_q{K}}{K}$ where $c_3$ is constant. (PROOF?)
%Therefore $1-\frac{M}{N}=\Theta(\frac{\log{K}}{K})$.
%Consider,
%\begin{align*}
%    F &\stackrel{(\ref{eqn35})}{\leq} Kq^{(k-t-m)m}q^{m^2}q^{2m}q \\
%    &\leq K K^{(k-t-m)} K^{m} K^{2} =K^{k-t+3} \\
%    &\leq K^{(\log_q{K}+3)} = q^{((\log_q{K})(\log_q{K}+3))}
%\end{align*}
%Therefore $F=q^{O(\log_q{K})^2}$.
%\section*{Acknowledgment}  This work was supported partly by the Early Career Research Award (ECR/2016/000447) from Science and Engineering Research Board (SERB) to Prasad Krishnan. Hari Hara Suthan C was supported by Visvesvaraya PhD Scheme for Electronics and IT.
%%%

\section{APPLICATION TO D2D NETWORKS}
\label{d2d}
We now adapt our new coded caching scheme to a scheme for D2D networks by utilizing a result of \cite{PDAapplications}. First we describe the D2D network model as in \cite{D2D} briefly. In contrast to the conventional coded caching setup, the central server is absent in D2D network. In a D2D coded caching network there is a library of $N^\mathcal{D}$ files, $K^\mathcal{D}$ users each equipped with a cache memory that can store $M^\mathcal{D}$ number of files. Each file is divided into $F^\mathcal{D}$ (subpacketization) number of equal sized subfiles. All users are connected by a bus link. During one time slot any one of the users can transmit and other users can receive (without error). The D2D coded caching system works in two phases. During the caching phase, the cache memory of each user is populated with contents available at the library (with the constraint that each cache can store $M^\mathcal{D}$ files). During the transmission phase each user demands any one of the files available in the library. The demand of each user is revealed to all other users. Every user $(l \in [K^\mathcal{D}])$ makes a multicast transmission of rate $r_l$ (the ratio of number of coded subfiles transmitted by $l$ to the subpacketization $F$), in its dedicated time slot to all other users using the bus link. From these multicast transmissions and cache contents each user decodes its demanded file. The rate of the D2D coded caching system is defined as $R^\mathcal{D} = \sum\limits_{l=1}^{K^\mathcal{D}}r_l$.

Similar to the conventional coded caching, the practical D2D coded caching systems demand low subpacketization schemes with lower rates.

% Similar to our coded caching scheme presented in Theorem \ref{main result}, the new D2D coded caching scheme (using \cite{PDAapplications}) achieves a subexponential subpacketization of $q^{O((log_qK)^2)}$ and rate $\Theta\left(\frac{K}{(log_qK)^2}\right)$, for large $K$, and the cached fraction $\frac{M}{N}$ being upper bounded by a constant $\frac{2}{q^{\alpha-1}}$ (for some prime power $q$ and constant $\alpha>1$).

% We begin with a definition for PDAs which is used to recall a result from \cite{PDAapplications}.
 
% In fact, it is not difficult to show that any $(c,d)-$ caching line graph is equivalently a  $d-(K,F,F-D,\frac{KD}{d})$ regular PDA also (and vice-versa).
In \cite{PDAapplications} it was shown that for any $g$-PDA with $g\geq 2$ there exists a corresponding D2D coded caching scheme (for the explicit construction, the reader is referred to Theorem 1 in \cite{PDAapplications}).

\begin{lemma}[Corollary $1$  in \cite{PDAapplications}]\label{D2D coded caching}
For a given $g-(K,F,Z,S)$ regular PDA with $g\geq 2$, there exists a scheme for a D2D network with $K^\mathcal{D}=K$ users and cached fraction $\frac{M^\mathcal{D}}{N^\mathcal{D}}=\frac{Z}{F}$, achieving the rate $R^\mathcal{D}=\frac{g}{g-1}\frac{S}{F}$, with subpacketization level $F^\mathcal{D}=(g-1)F$. (Here the parameters with superscript $\mathcal{D}$ represents the parameters of the D2D coded caching scheme)
\end{lemma}

% \begin{corollary}
% For a D2D system with $K$ users, where $K$ is a non-prime integer, then
% \begin{itemize}
%     \item For $M\in \{\frac{tN}{K}: t|K, 1<t<K$\}, there exists a caching scheme achieving rate $R_1=\frac{1-M/N}{M/N-1/K}$,
%     with subpacketization level $I_1=\frac{M}{N}(\frac{KM}{N}-1) (\frac{N}{M})^{\frac{KM}{N}}$.
    
%     \item For $M\in \{\frac{tN}{K}: (K-t)|K, 1<t<K$\}, there exists a caching scheme achieving rate $R_2=\frac{1-M/N}{M/N-1/K}$,
%     with subpacketization level $I_2=\frac{M}{N}(\frac{KM}{N}-1) (\frac{N}{N-M})^{K(1-\frac{M}{N})}$.
% \end{itemize}
% \end{corollary}

% Therefore, our scheme also results in a D2D transmission scheme with parameters appropriately as mentioned above. Further, as the subpacketization parameter for the D2D scheme is only dependent on the subpacketization of the underlying coded caching scheme, any coded caching scheme which has a reduced subpacketization (like the present scheme) will result in a reduced subpacketization for the D2D scheme also. Hence, our coded caching scheme, when adapted as a D2D schemes achieves a small subpacketization compared to prior known schemes obtained from other coded caching schemes. 

By Lemma \ref{line graph pda connection} it is easy to see that the $(c,d)$-caching line graph developed in Section \ref{our scheme B subsection} corresponds to a $d-(K,F,F-D,\frac{KD}{d})$ regular PDA.  Now, by applying Lemma \ref{D2D coded caching}, we can get the corresponding D2D coded caching scheme which is presented in the Theorem \ref{Our D2D scheme} (the proof follows from Lemma \ref{line graph pda connection} and Lemma \ref{D2D coded caching}).

% We now present a new D2D coded caching scheme by applying Theorem \ref{main result} to Lemma \ref{D2D coded caching}.
\begin{theorem}\label{Our D2D scheme}
The caching line graph given in Section \ref{our scheme B subsection} corresponds to a D2D coded caching scheme with 

\[K^\mathcal{D}=\frac{q}{2}\gbinom{k-t+1}{1} \gbinom{k-t}{1}.\] 
\[F^\mathcal{D}=\frac{(m+1)(m+4)}{2} \gbinom{k-t+1}{m+1}\dfrac{\prod\limits_{i=0}^{m}(q^{m+1}-q^{i})}{(m+1)! (q-1)^{m+1}}.\] 

\[\frac{M^\mathcal{D}}{N^\mathcal{D}}=1-q^{2(m+1)}\dfrac{\gbinom{k-m-t}{1}\gbinom{k-m-t-1}{1}}{{\gbinom{k-t+1}{1}\gbinom{k-t}{1}}}.\] 
 
\[R^\mathcal{D}=\dfrac{q^{(2m+3)}}{(m+1)(m+4)}\gbinom{k-m-t}{1}\gbinom{k-m-t-1}{1}.\]
\end{theorem}

By following the similar techniques as in Section \ref{asymptotics}, it is not difficult to show that $F^\mathcal{D}=q^{O((\log_q{K^\mathcal{D}})^2)}$ and $R^\mathcal{D}=\Theta(\frac{K^\mathcal{D}}{(\log_q{K^\mathcal{D}})^2})$ (where $K^\mathcal{D}$ is the number of users in the D2D coded caching scheme). 

In Table \ref{tab2}, we compare the D2D coded caching scheme presented in Theorem \ref{Our D2D scheme} with that of \cite{hypercubeD2D,D2D}.
As far as possible, we choose corresponding values for $K^\mathcal{D}, U^\mathcal{D}=1-\frac{M^\mathcal{D}}{N^\mathcal{D}}$.
The parameters corresponding to Theorem \ref{Our D2D scheme} are labelled as $K_1^\mathcal{D}, U_1^\mathcal{D}, F_1^\mathcal{D}, R_1^\mathcal{D}$. 
The parameters corresponding to the scheme of \cite{hypercubeD2D} are
$K_2^\mathcal{D}, U_2^\mathcal{D}=1-\frac{1}{y_1}, F_2^\mathcal{D}=y_1^{y_1}, R_2^\mathcal{D}=y_1$ where $y_1=\sqrt{K^\mathcal{D}}$. In this scheme we only have freedom to choose $K_2^\mathcal{D}$, all other parameters depend on $K_2^\mathcal{D}$ (because of this we are unable to even approximately match $U_2^\mathcal{D}$ with $U_1^\mathcal{D}$ of our scheme).

The parameters corresponding to the scheme of \cite{D2D} are $K_3^\mathcal{D}, U_3^\mathcal{D}=1-\frac{M^\mathcal{D}}{N^\mathcal{D}}, F_3^\mathcal{D}=y_2\binom{K^\mathcal{D}}{y_2}, R_3^\mathcal{D}=\frac{N^\mathcal{D}}{M^\mathcal{D}}-1$ where $y_2=\lfloor \frac{M^\mathcal{D}K^\mathcal{D}}{N^\mathcal{D}}\rfloor$. From Table \ref{tab2} it is clear that the scheme presented in Theorem \ref{Our D2D scheme} performs much better than the schemes of \cite{D2D,hypercubeD2D} in terms of subpacketization but with higher rate, which indicates that our scheme can be implemented in the practical D2D coded caching networks.
\begin{remark}
At the time that we were finalizing this paper, we became aware of a recent work \cite{D2DPDAJan2019} on D2D schemes with low subpacketiation based on PDAs. While we are yet to do a rigorous comparison, superficial observations suggest that our scheme will continue to retain its advantages over those in \cite{D2DPDAJan2019}.
\end{remark}

\bibliographystyle{IEEEtran}
\bibliography{IEEEabrv,cite.bbl}

\end{document}